\documentclass[12pt,preprint]{aastex}

\shorttitle{Nonresonant instability}
\shortauthors{Ohira et al}

\begin{document}

\title{Two-Dimensional particle-in-cell simulations of the nonresonant, cosmic-ray driven instability in SNR shocks}

\author{Yutaka Ohira\altaffilmark{1}, Brian Reville\altaffilmark{2}, 
John G. Kirk\altaffilmark{2} and Fumio Takahara\altaffilmark{1}}

\begin{abstract}

In supernova remnants,
the nonlinear amplification of magnetic fields upstream of
collisionless shocks is essential for the acceleration 
of cosmic rays to the energy of the \lq\lq knee\rq\rq\ at
$10^{15.5}\,$eV.  A nonresonant instability driven by the cosmic ray
current is thought to be responsible for this effect. We perform
two-dimensional, particle-in-cell simulations of this instability.  We
observe an initial growth of circularly polarized non-propagating
magnetic waves as predicted in linear theory. 
It is demonstrated that in some cases 
the magnetic energy density in the growing waves,
can grow to at least 10 times its initial value. We find no
evidence of competing modes, nor
of significant modification by thermal effects. At late times we observe saturation
of the instability in the simulation, but the mechanism responsible is an
artefact of the periodic boundary conditions and has no
counterpart in the supernova-shock scenario.
\end{abstract}

\keywords{supernova remnants -- shock waves -- plasmas -- 
cosmic rays}

\altaffiltext{1}{Department of Earth and Space Science, 
Graduate School of Science, Osaka University, 1-1 Machikaneyama-cho, 
Toyonaka, Osaka 560-0043, Japan; yutaka@vega.ess.sci.osaka-u.ac.jp}
\altaffiltext{2}{Max-Planck-Institut f{\"u}r Kernphysik, Heidelberg 69029, Germany}

\section{Introduction}

Diffusive shock acceleration (DSA) at supernova remnant shocks
is widely considered to be the primary source of galactic cosmic rays
\citep[for a recent review see][]{Hillas05}. A crucial aspect of
DSA is the self-excitation of hydromagnetic waves 
due to currents produced by the streaming energetic ions.
These provide the pitch-angle scattering necessary to allow spatial 
diffusion of the relativistic particles. 
The maximum 
acceleration rate 
that can be obtained corresponds to the smallest possible spatial diffusion 
coefficient. In this \lq Bohm limit\rq\ 
the maximum particle energy is 
determined either by the age of the remnant \citep{lag83},
or by its geometry \citep{Berezhko}. Even adopting optimistic
values for the upstream parameters of a supernova remnant shock,
particles are limited to energies below the knee of the
cosmic-ray spectrum at $\sim 10^{15.5}$~eV. However, 
nonlinear amplification of the upstream magnetic field due
to cosmic-ray currents may provide a possible solution to
this problem \citep{Quest, luc00}.

There is increasing observational evidence that the magnetic fields
immediately downstream of several young supernova remnant shocks are
much stronger than would be obtained by compressing the ambient
interstellar field at an MHD shock front \citep{Vink,ber03,bam05,uch07}. 
These observations give additional motivation to the
development of theoretical models of nonlinear field generation by
both plasma instabilities in the precursor~\citep{luc00,bel04} and
fluid-type instabilities in the downstream plasma~\citep{gia07}.
These two mechanisms can, in principle, operate
simultaneously \citep{zir08}. However, to reach the highest energies, particles
must be speedily returned to the shock from excursions into both the
upstream and downstream plasmas. The generation of an
amplified magnetic field 
in the precursor that is subsequently advected into the downstream
region is sufficient to ensure this, whereas amplification of only the 
downstream field is not. 

Using a hybrid kinetic-MHD analysis, it was demonstrated by \cite{bel04} 
that efficient cosmic-ray acceleration can, in the linear regime,
drive a short wavelength, almost aperiodic, instability with wave-vector
parallel to the ambient field. The growth of this mode
can be considerably more rapid than that of the resonantly 
driven modes previously considered \citep{mcken,acht83}. 
The instability is driven by the reaction of the thermal plasma, as
it attempts to compensate the current produced by the streaming cosmic rays. 
Numerical investigations of this instability have 
been performed using both MHD \citep{bel04, bel05, rev08, zpv08} 
and Particle-In-Cell (PIC) simulations \citep{nie08, riq08}. 
While the results of MHD simulations show that the perturbed 
magnetic field $\delta B$ becomes much larger than the initial 
uniform magnetic field $B_0$, the PIC simulations of 
\cite{nie08} did not observe the 
parallel mode predicted in the linear analysis, and only moderate 
amplification of the magnetic field was observed.
On the other hand, the recent results reported by
\citet{riq08} do identify the parallel mode and propose
an explanation of its saturation in the non-linear stage.

In this paper we report new PIC simulations of this instability.
We extend the range of parameters beyond that 
investigated by \cite{nie08}, and select values that more 
closely represent the environment in an SNR precursor. Although both 2D and 3D
simulations in a similar parameter range have been performed
\citep{riq08}, we restrict our work to 2D. Our results
--- obtained independently and with some technical differences in the
numerical treatment --- can be regarded as complementary, and our
physical interpretation, at least of the early stages of the non-linear
development, are congruent. In section~2, we recall the 
linear kinetic analysis 
of the nonresonant instability, in order 
to elucidate the conditions 
that must be satisfied in the PIC simulations. 
The details of our simulation 
parameters are described in section~3. 
We present the results of the simulations in 
section~4 followed by a discussion of the 
saturation and its implications in section~5.
We conclude with a summary of the results.

\section{Linear analysis}
\label{lineartheory}

We review the linear kinetic analysis of an 
electron-ion plasma
upstream of a non-relativistic shock with a power-law 
distribution of streaming cosmic rays.  We focus our
attention on parallel shocks, with the streaming velocity directed along the
zeroth order magnetic field. The kinetic linear dispersion relation 
for circularly polarized hydromagnetic waves propagating
parallel to the mean magnetic field in such a plasma was investigated
by \cite{acht83}, subject to the conditions of zero net charge and current.
In the upstream plasma frame the cosmic ray streaming velocity is 
approximately that of the shock velocity \citep{McClements96}.
Using the diffusion approximation it can be shown
that, for a power-law momentum distribution of streaming
cosmic rays with a spectral index $s>3$ $(f_{\rm cr}(p)\propto p^{-s})$, the dispersion 
relation can be written as \citep{rev07}:
\begin{eqnarray}
\label{disp}
{\omega}^2
+\epsilon \left(\frac{k^2V^2_{\rm ti}}{\Omega_{\rm ci}}\right)
{\omega}
-v_{\rm A}^2 k^2-
\epsilon \zeta v_{\rm s}^2 \frac{k}{r_{\rm g}}
\left(\sigma(kr_{\rm g})-1\right)=0,
\end{eqnarray}
where $\omega=\omega_r+{\rm i}\gamma$ is the frequency of the waves in the upstream 
ion rest frame, $k>0$ the wavenumber, $v_{\rm A}$ the Alfv\'en velocity,
$V_{\rm ti}$ and $\omega_{\rm ci}$ the ion thermal velocity 
and cyclotron frequency, $v_{\rm s}$ the speed of the shock and 
$r_g=p_{\rm min}c/eB_0$
the gyroradius of the lowest energy cosmic ray. The polarization
of the waves is determined by $\epsilon = +1(-1)$ for right (left) handed
modes ($\omega_r > 0$). This result agrees with that found by \citet{bel04} using an MHD approach,
and extends it by including an additional term representing thermal ion damping.
The growth rate of the unstable modes is determined by the driving
parameter
\begin{equation}
\zeta = 
\frac{n_{\rm cr} p_{\rm min}}{n_{\rm i} m_{\rm i} v_{\rm s}},
\nonumber
\end{equation} 
where $n_{\rm cr,i}$ are the densities of cosmic rays and ions.
Finally $\sigma$ is a complex function
describing the electric fields produced by the cosmic ray current. It
can be shown that $\sigma$ is a decreasing function of 
wavenumber for $kr_{\rm g}>1$ \citep{rev07}. 
In particular, for wavelengths much less than
the gyroradius of the cosmic rays, $\sigma$ can be neglected with respect to 
unity, and one finds that there exists a purely growing mode with 
growth rate
\begin{equation}
\label{gmax}
 \gamma_{\rm NR} = \frac{\zeta}{2}\frac{v_{\rm s}}{v_{\rm A}}
\frac{v_{\rm s}}{r_{\rm g}}
\approx\frac{1}{2}\frac{v_{\rm s}}{v_{\rm A}}
\frac{n_{\rm cr}}{n_{\rm i}}\Omega_{\rm ci}.
\end{equation}
An important constraint in deriving Eq.~(\ref{disp}), is the
magnetization condition, $|\omega| \ll \Omega_{\rm ci}$. 
On length scales much shorter than their gyroradius,
the cosmic rays are essentially unmagnetized, i.e., their trajectories 
are rectilinear. The fact that the background plasma is magnetized leads
to an asymmetry in the system, resulting in an uncompensated perpendicular 
current. This is what drives the growth or decay of all 
elliptical modes with ${\bf k \cdot B_0} \neq 0$ \citep{bel05, melrose05}.
If the background ions are unmagnetized, there is no inertia to
support the growing waves, and the ions quickly act to compensate the
cosmic-ray current. Significant heating of the plasma causes the
ions to behave as if they were unmagnetized, thereby 
reducing the growth-rate. The same effect arises 
in the relativistic analysis of this instability
\citep{rev06}. 

In order to maintain a steady galactic cosmic-ray energy density,
supernova remnants must channel approximately 10~percent of the
kinetic energy entering the shock front into cosmic rays.  Assuming a
shock velocity of $v_{\rm s}=0.01c$ with a cosmic-ray distribution
$f_{\rm cr}(E)\propto E^{-2} $ from $10^7$~eV to $10^{15}$~eV, 
this implies a cosmic-ray number density of 
$n_{\rm cr}=10^{-5}\,{\rm cm}^{-3}$.  Taking fiducial interstellar medium values for the
ion number density and magnetic field: $n_{\rm i}=1\,{\rm cm}^{-3}$
and $B=3~\mu {\rm G}$, the maximum growth rate of the nonresonant
instability is 
$\gamma_{\rm NR}\sim 2.5\times 10^{-3} {\Omega_{\rm ci}}$.  
In the upstream rest frame, the cosmic rays drift with
respect to the thermal ions at the shock velocity $v_{\rm s}$ . To
compensate this current, there is relative velocity between upstream
ions and electrons, $V_{\rm d}=v_{\rm s}n_{\rm cr}/n_{\rm
  e}=10^{-7}c$. This value is much smaller than the ion and
electron thermal velocity. Hence an MHD description of the plasma is
appropriate for the parameters in the precursor of a supernova
remnant.  P.I.C. simulations of this essentially MHD instability are
computationally very intensive.  However, they are the only way of
quantifying the relative importance of high frequency waves and other
kinetic effects on the nonlinear evolution of the instability.

\section{Simulation}
\label{simulation}

We use a 2D fully relativistic electromagnetic PIC code 
with a
fast algorithm that solves for the current density and conserves
charge \citep{ume03}. The
simulations are carried out in the upstream rest frame.  While the
simulations of \cite{nie08} used a full kinetic treatment of the
cosmic rays, during the linear stage of growth there was no
significant reduction in the streaming energy of the cosmic rays.  In
order to identify the growth and evolution of the instability, we
maintain a constant, uniform external cosmic-ray current throughout
the simulations, as has been used in previous MHD
simulations \citep{bel04, rev08}. 

Similar to previous simulations, we investigate the role 
played by streaming cosmic rays of uniform density in a homogeneous
electron-ion plasma. This is intended to represent a small region
in the precursor of a supernova remnant shock. However, when, during
the course of the simulation, the magnetic field fluctuations
associated with the nonresonant instability become strong, they begin
to damp the relative drift velocities of the cosmic rays and the gas.
In a shock precursor, a change in relative drift velocity is
automatically associated with an increase in the gas and cosmic ray
densities, as the shock is approached. The simulation box, however,
has periodic boundary conditions and, therefore, simulates a spatially
uniform plasma.  These boundary conditions therefore fail to reproduce
the behavior expected in the shock precursor, as soon as the relative
drift velocity changes significantly. For this reason, the saturation
of the instability observed in the simulation, which occurs when the
drift speeds become equal, does not correspond to the saturation
mechanism expected in the shock precursor. However, the initial linear
phase of the instability and a substantial part of its nonlinear
evolution are accurately modelled.

\subsection{Setting}

We define the zeroth order magnetic field to be along the 
negative $x$-direction, and the 
cosmic rays stream in the positive $x$-direction. Periodic boundary 
conditions are used in both the $x$ and $y$ directions. 
The electrons, ions and cosmic rays are initialized
such that the overall 
charge and current densities vanish. Thus, 
$n_{\rm e} = n_{\rm i} + n_{\rm cr}$
and the electrons have a
net drift velocity in the $x$-direction of $V_{\rm d,e}=V_{\rm d,cr}
n_{\rm cr}/n_{\rm e}$. (Subscripts e, i and cr represent electrons, 
ions and cosmic rays, respectively.) At the beginning of the simulation,
each population is uniformly distributed in the $x-y$ plane with a
Maxwellian momentum distribution at equal temperatures 
${ T=T_{\rm e}=T_{\rm i}}=1.3{\rm~keV}$. 
The length of each cell $\Delta x = \Delta y$ and time step $\Delta t$ 
of the simulation are twice the Debye length and 
$0.0714\omega_{\rm pe}^{-1}$, respectively.  
Initially, each cell contains 49 positively and 49 negatively
charged macroparticles. In the presence of a uniform cosmic ray
density, the charge of each species of macroparticle 
must be chosen such that the overall plasma is 
initially charge neutral. The charge to mass ratio of the negatively 
charged macroparticles corresponds to that of an electron.

The values for $\Omega_{\rm ce}/ \omega_{\rm pe}$, $n_{\rm cr}/n_{\rm i}$, 
$V_{\rm d}/c$ and the number of cells for each simulation run are 
given in Table~\ref{table1}, where $\Omega_{\rm ce}$ and $\omega_{\rm pe}$ 
are the electron cyclotron frequency and electron plasma frequency, 
respectively. We also specify the mass ratio
$m_{\rm i}/m_{\rm e}$ for each run. To make the problem more tractable, 
we are forced to use unrealistically high values for the cosmic-ray
number densities. This is done purely to reduce the computation time, 
and, as we show, the essential physical mechanisms are still well-captured.

The size of the simulation box in the $x$ and $y$-directions for run A
is taken to be $L_x=L_y=6.66\lambda_{\rm NR} $, 
where $\lambda_{\rm NR}=2\pi v_{\rm A}/\gamma_{\rm NR}^{\rm max}$ is 
the wavelength of the most unstable mode of the nonresonant instability. 
The simulation is followed until 40$\tau_{\rm grow}$, 
where $\tau_{\rm grow}=\gamma_{\rm NR}^{-1}$. 
The parameters used in run B are chosen such that the magnetization condition 
$\gamma_{\rm NR} < \Omega_{\rm ci}$ is not satisfied. As discussed in section
\ref{lineartheory}, it is not expected that the nonresonant current-driven 
instability will operate in such a situation. Similar parameters
have been used by \cite{nie08}, and we perform this simulation in order to 
facilitate comparisons with that work.
The size of the simulation box for run-B is $L_x=L_y=9.65\lambda_{\rm NR}$.

We also perform two additional simulations, runs C and D, with higher 
cosmic-ray drift velocity and mass ratio, respectively. The box size is the 
same as that used in run A, $L_x=L_y=6.66\lambda_{\rm NR} $.
While the conditions
in run C are less appropriate for a supernova shock precursor than those of
run A, the increased cosmic-ray drift speed provides a greater amount of free 
energy in the system. Likewise, the magnetization condition in run A is satisfied
better than in run D, however, the larger mass ratio allows us to investigate the effect
that increased inertia has on the thermal ions.

\section{Results}
\label{simresults}

The evolution of the spatially averaged energy density of the magnetic
and electric field components associated with the growing waves, 
for run-A is shown in Figure 1.
The linear growth rate of the spatially averaged transverse component
of the magnetic field is approximately $\sim0.5\gamma_{\rm NR}$. The
characteristic wave length is almost the same as that predicted by the
linear analysis.  Figure 2a shows the spatial distribution of the
$z-$component of magnetic field during the linear phase of growth
($t=14\tau_{\rm grow}$)\footnote{Note $t>\tau_{\rm grow}$ does not
  mean nonlinear phase because the initial perturbation is very
  small.}  , where the $k$-vector is clearly directed along the zeroth
order field.  The $y$-component of the magnetic field has a similar
structure with the phase of the wave pattern shifted by $\pi/2$.

For comparison, we show in Figure 2b, the $z-$component of the magnetic
field during the linear phase of growth for run-B. The ions in this 
case are not magnetized, and the nonresonant mode is
not observed. The wave is aperiodic and its wavevector is 
almost perpendicular
to the beam. This is typical of Weibel-type instabilities. We discuss this
further in the next section.

The linear development of the fluid quantities in the simulations is similar to
that of previous MHD simulations \cite{bel04, bel05, rev08, zpv08}.
The initial stage sees the development of the aperiodic instability, with 
uniform density. As the fastest growing mode emerges from the initial noise, 
the net ${\bf j \times B}$ force begins to push the plasma in the direction 
transverse that of the cosmic ray drift, generating low
density regions between filaments of compressed plasma. 
The filaments are uncorrelated in the direction of
the cosmic ray drift, similar to what is observed in MHD simulations 
\citep[e.g.][]{bel04, rev08}. The growth of the cavities 
is eventually limited due to the expansion of neighbouring cavities,
at which point they appear to have a radius of $\lambda_{\rm NR}$.

The late-time nonlinear evolution, however, differs from that of previous MHD 
simulations. While both see the eventual disruption of the filamentary structures,
the particle in cell simulations see an acceleration of the background plasma in 
the direction of the cosmic-ray drift. 
This effect is not observed in simulations that
represent the system as a single MHD fluid driven by a fixed external
current, because the charge of the streaming particles, and
hence, the direction in which they drift, is undefined.
Figures 3(a) and 3(b) show the distribution of the total magnetic field energy and upstream 
proton density at the end of the simulation ($t=40\tau_{\rm grow}$).
Locally, the amplification factor $\delta B/ B_0$ is about~16 but the 
global, spatially averaged value is about~4. 
These images differ quite considerably from the nonlinear results presented
in previous MHD simulations \citep{bel04, zpv08}, not only because of the net drift,
but also due to the anti-correlation between the magnetic field and the plasma density 
(Figure 3(b)). 

At the point at which the magnetic field ceases to grow, 
the bulk plasma is drifting with a constant 
velocity $v_{d,i}=0.9v_{d,cr}=0.09c$. This is almost the shock velocity.
The development of run~C and run~D are similar to that of run~A. However, the
saturated field values are slightly larger, since the initial amount of free energy
available is larger in both cases. The growth of the different 
field components is plotted for run C and run~D in Figures 4 and 5, respectively.
We discuss the physical nature of the field saturation in the next section.
  
\section{Discussion}
\label{discussion}

We find that the nonresonant cosmic-ray driven instability develops in
all cases where the background plasma is magnetized. 
From Figure \ref{fig2}(a) and \ref{fig2}(b), it can be seen that there are 
considerable differences between the results of run~A and run~B although the observed wavelength is similar to $\lambda_{\rm NR}$. 
The most unstable mode in run~B is the Weibel type
instability, occurring between the counterstreaming electrons and 
ions. Neglecting thermal effects and 
the magnetic field, the growth rate $\gamma_{\rm WI}^{\rm max}$ and the 
wavelength $\lambda_{\rm WI}$ of the Weibel instability are
\begin{equation}
\gamma_{\rm WI}^{\rm max}=\frac{n_{\rm cr}}{n_{\rm e}}\frac{v_{\rm s}}{c}\omega_{\rm pi}, \ \ \lambda_{\rm WI}=2\pi \frac{c}{\omega_{\rm pe}},
\end{equation}
where the direction of the wave vector is perpendicular to the drift direction.
Comparing the two theoretical growth rates and the wavelengths we see that
\begin{equation} 
\frac{\gamma_{\rm WI}^{\rm max}}{\gamma_{\rm NR}^{\rm max}} = \frac{n_{\rm e}}{n_{\rm i}} \simeq 1, \ \ \frac{\lambda_{\rm WI}}{\lambda_{NR}}=\frac{1}{2}\frac{n_{\rm cr}}{n_{\rm i}}\frac{V_{\rm d,cr}}{c}\frac{\omega_{\rm pe}}{\Omega_{ce}} \simeq 1. 
\end{equation}
For the choice of parameters used in run~B,  the wavelengths of
the different instabilities are quite similar,
even though the mechanism is quite different.
This is consistent with the results of \citet{nie08},
and we also find the fastest growing mode to be slightly oblique.
Hence we may be observing a mixed-mode between the Weibel 
and the nonresonant current driven instability or some other type 
of mixed-mode instability similar to those discussed in \cite{Bret}.

In the late stages of the simulations we see the emergence of a
non-thermal power law tail in the ion distribution.   
The use of a constant external cosmic-ray current
prevents the development of the Buneman instability, a source
of ion heating in the linear development. This dramatically 
reduces the heating in the initial stages, that was previously observed
in \cite{nie08}. Figure 6 shows
the final electron and ion energy distribution for run~A, C and
D. The distributions are essentially isotropic, and since the
bulk drift velocity is small in comparision with the thermal velocities,
we can safely calculate the spectrum in the box frame.
The electron energy distribution can be well fitted with a
Maxwellian distribution with temperature $10$ keV for run~A and $40$
keV for run~C and D. A non-thermal tail appears in the distribution of
the ions with a very soft power-law index of $\sim 7.3$ for Run-A,
but a relatively hard $\sim2$ for run~C and D. The 
mechanism responsible for the formation of this 
power-law distribution is not clear, and although our simulations
have a relatively large number of particles per cell, we cannot rule out a 
numerical artefact. As
regions of oppositely polarized magnetic field collide each other, an
anti-parallel configuration of the magnetic fieldlines appears.  It is,
therefore, 
possible that collisionless magnetic dissipation may also play an
important role in plasma heating mechanism and the saturation of the
field growth as well as the reduction in the drift velocity between
the electrons and ions. 
This is an interesting process in its own right,
and further investigation is warranted. 
However, in the current context, it is not clear what
role these mechanisms play, since the late-stage nonlinear behavior 
has no counterpart in the supernova-shock scenario.

In agreement with \cite{nie08} and \cite{riq08}, 
we observe in our simulations that the
magnetic field growth ceases when the plasma bulk velocity becomes
comparable to the cosmic-ray drift velocity. Although we have fixed
the cosmic ray current in our simulations, the observed saturation is
clearly due to the reduction in the relative speeds of plasma and cosmic rays.
Indeed, if the cosmic rays are isotropic in the plasma frame, no streaming
instability can operate. 
The simulations are performed in a box with periodic boundary 
conditions in space. The plasma speed is initially zero, 
and the cosmic ray streaming speed is maintained
constant in time and space. It represents the speed with which the shock
front approaches the simulation box.   
Within this picture, the instability can saturate only when the entire
plasma moves with the drift speed of the cosmic rays. 

In the precursor of a shock front, conditions are different. 
There, the interaction 
between plasma and cosmic rays defines the diffusion length scale of the 
cosmic rays. This is the scale on which the cosmic ray current, or,
equivalently, pressure, decays with distance ahead of the shock front. 
In the case of efficient acceleration, it is also the length scale on
which the precursor plasma is compressed.
Simulations can model this situation provided the box size is small
compared to the diffusion length scale, and the time scales are short
compared to the time on which the cosmic ray current and 
plasma density change because of the approaching shock front.
However, this restriction means that the simulations no longer model 
a precursor when they are followed until the plasma speed approaches the 
shock speed (which equals the cosmic-ray drift speed).  

This can be seen explicitly from the equation of 
conservation of mass in a stationary precursor,
$\rho(x) u(x) = \mbox{constant}$, which dictates that any change in the flow 
velocity must be associated with a change in the density of the fluid.
In the present context,
the simulation box is in the upstream plasma frame,
approaching the shock with velocity $v_{\rm s}$. 
Defining the density and bulk velocity inside the box to be $\rho_b$ and
$u_{\rm b}$ respectively, we see that
\begin{equation}
 \rho_{\rm b}= \frac{v_{\rm s}\rho_0}{v_{\rm s}-u_{\rm b}}
\end{equation}
where $\rho_0$ is the initial density of the thermal plasma. Thus, as
the flow speed inside the box increases, the density should also
increase. Since the total density in the box is fixed throughout the
simulation, the observed growth rate is only accurate provided the
bulk velocity inside the box is small compared to the shock velocity,
or, equivalently, the streaming speed of the cosmic rays, i.e.,
$u_{\rm b} \ll v_{\rm s}$. This corresponds to $\tau_{\rm
  grow}< 20$ for run A and $\tau_{\rm grow}< 23$ for runs C and D. 
The amplification of the energy density in
the total field in run~A is not yet substantial, but in runs~B and C 
roughly 10 times the initial energy density is reached. The final 
saturation levels are much higher, but our simulations do not 
necessarily imply that these can be reached in a precursor scenario, 
and a PIC simulation of this
case is currently out of reach.

\section{Summary}

We have performed 2D PIC simulation to investigate the nonlinear
physics of the nonresonant current instability. We have demonstrated
that the energy in the growing waves can substantially exceed that of
the initial seed field. However, limitations inherent in the simulation
method lead to an artificial saturation level. 
Although we cannot compute a realisitic saturated
value of the magnetic field, substantial amplification is seen
in some runs before they lose validity, 
and the possibility remains open that the field would
continue to grow, transferring the energy to
longer lengthscales, in accordance with theoretical predictions. 
Thus, as predicted, magnetic field amplification is triggered by
the nonresonant streaming instability and is likely
to play an important role in the 
acceleration of cosmic rays to the knee and beyond in supernova remnants.

\acknowledgments Y.O. is grateful to S. Matsukiyo and T. Umeda for
discussions on the nonresonant instability and PIC
simulations. B.R. thanks A. Spitkovsky for many useful discussions.
Y.O. is supported by a Grant-in-Aid for
JSPS Research Fellowships for Young Scientists.  BR gratefully
acknowledges support from the Alexander von Humboldt foundation.
Numerical computations were carried out on the Cray XT4 at Center for
computational Astrophysics, CfCA, of the National Astronomical Observatory
of Japan and on the Blue Gene/P at the Rechenzentrum Garching of the Max
Planck Society.

\clearpage

\begin{figure}
\plottwo{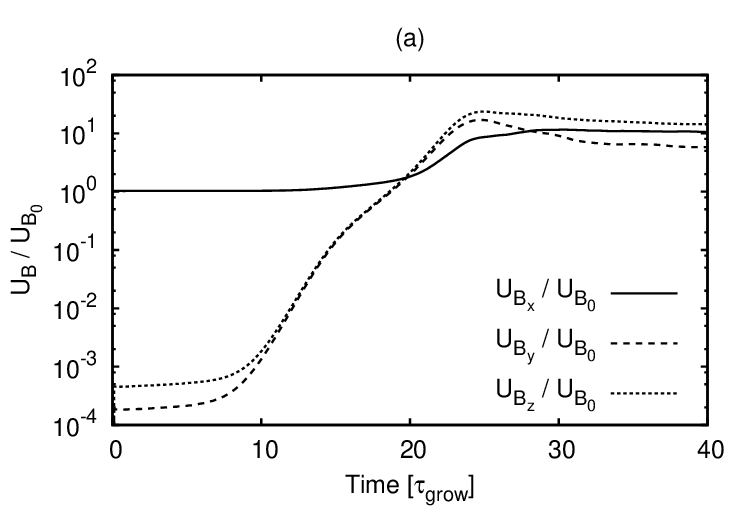}{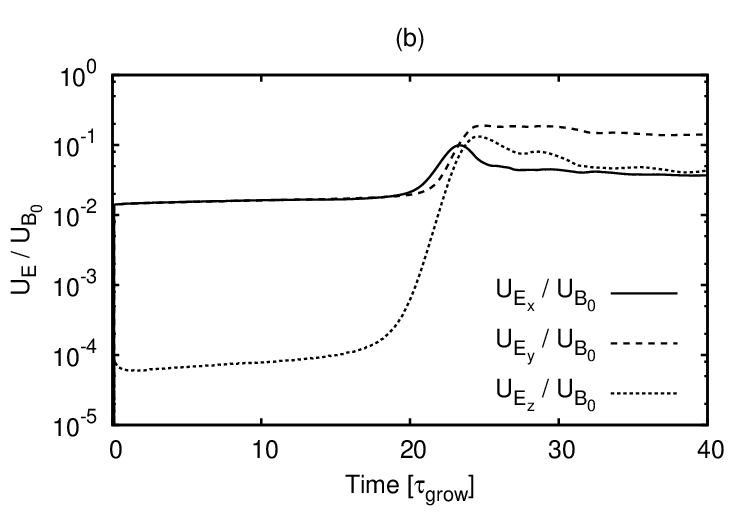} 
\caption{Time development of the spatially averaged root mean square of (a) magnetic field, (b) electric field for Run-A. Solid, dashed and dotted curves represent the $x$, $y$ and $z$-components, respectively.
\label{fig1}}
\end{figure}

\clearpage

\begin{figure}
\plottwo{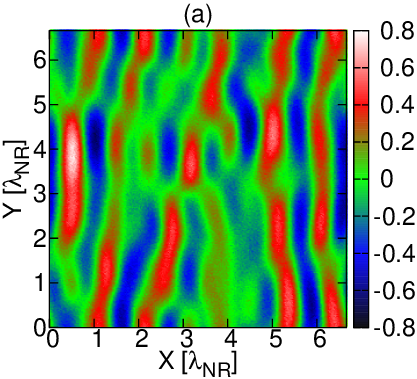}{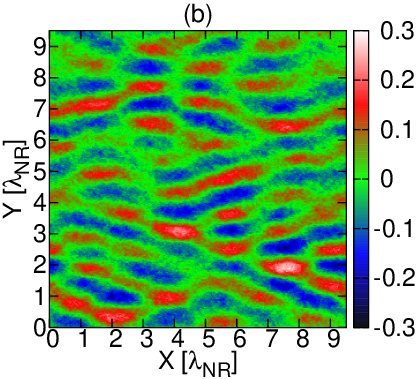}
\caption{$z$-component of magnetic field at (a) $t=14\tau_{\rm grow}$ in run-A, (b) $t=8 \tau_{\rm grow}$ in run-B.
\label{fig2}}
\end{figure}

\clearpage

\begin{figure}
\plottwo{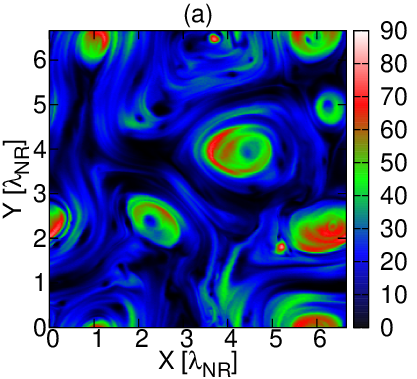}{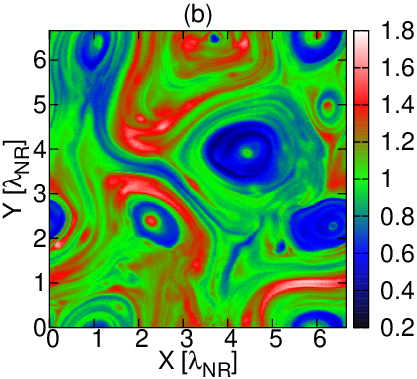}
\caption{(a) The magnetic field energy density, and (b) density of ions at $t=40\tau_{\rm grow}$ in run-A \label{fig3}}
\end{figure}

\clearpage

\begin{figure}
\plottwo{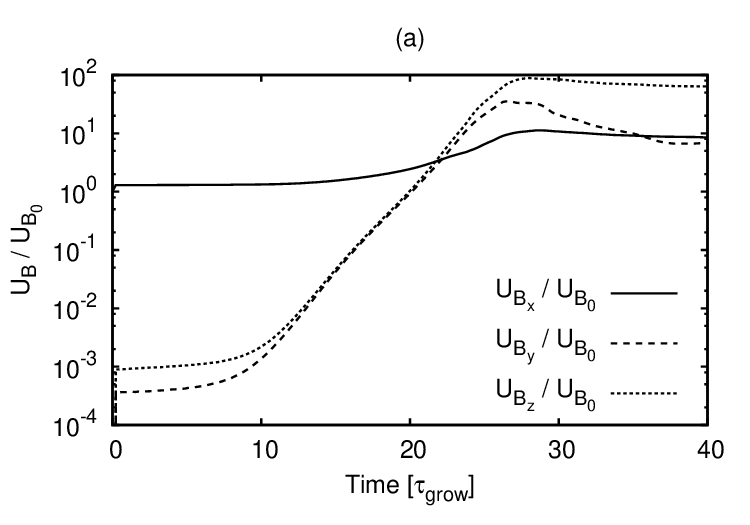}{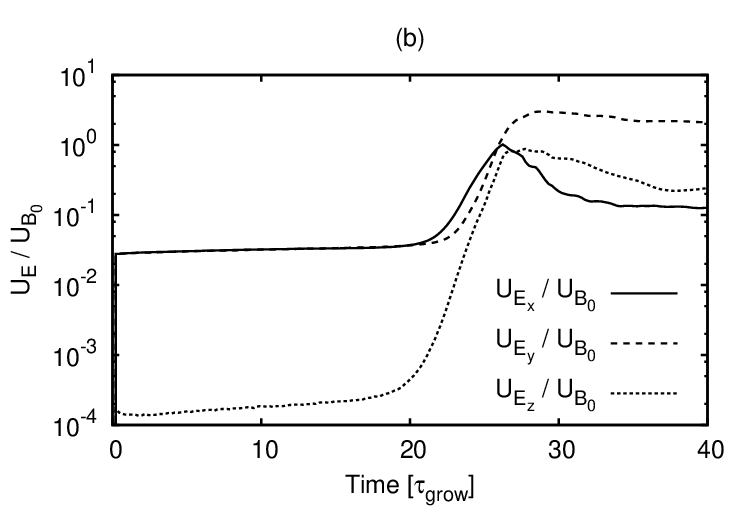} 
\caption{Time development of the spatially averaged root mean square of (a) magnetic field, and (b) electric field for Run-C. Solid, dashed and dotted curves represent the $x$, $y$ and $z$-components, respectively.
\label{fig4}}
\end{figure}

\clearpage

\begin{figure}
\plottwo{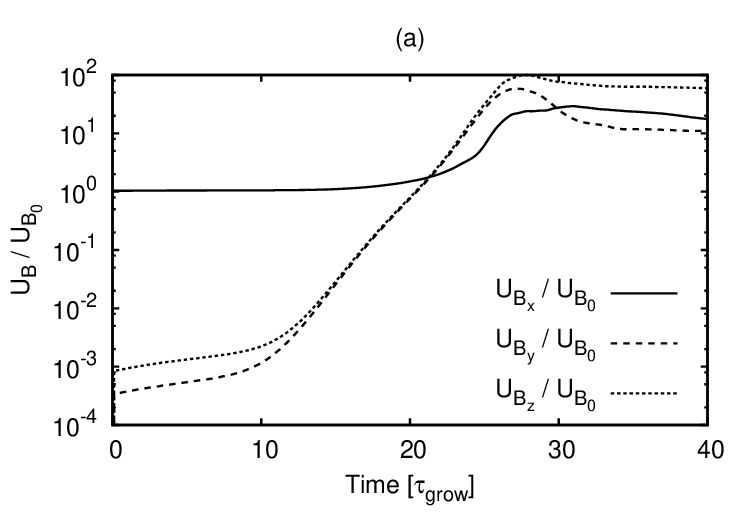}{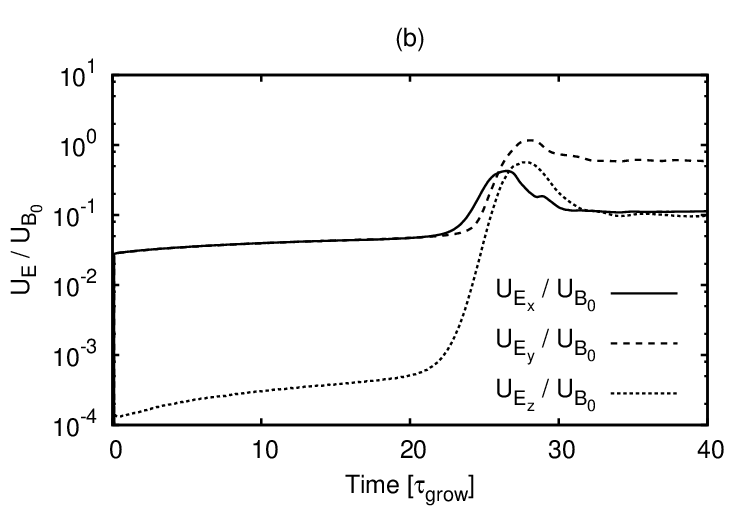} 
\caption{Time development of the spatially averaged root mean square of (a) magnetic field, and (b) electric field for Run-D. Solid, dashed and dotted curves represent the $x$, $y$ and $z$-components, respectively.
\label{fig5}}
\end{figure}

\clearpage

\begin{figure}
\plotone{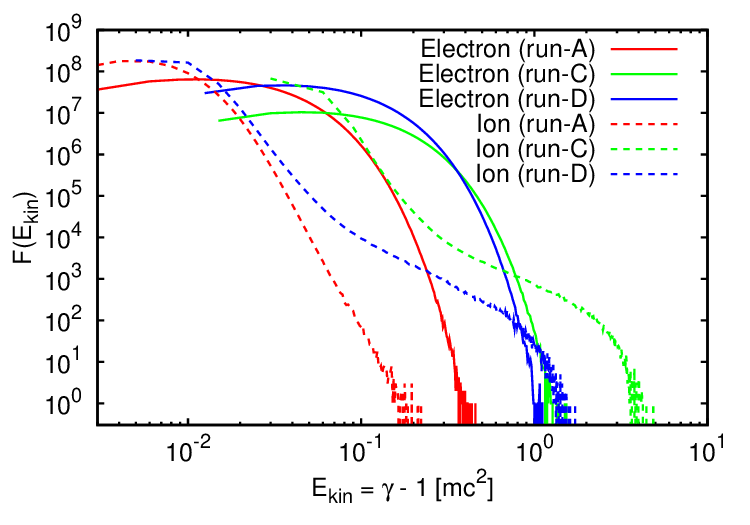}
\caption{Energy distribution at the final stage in run-A, C and D \label{fig6}}
\end{figure}

\clearpage
\begin{deluxetable}{lccccc}
\tablecaption{Simulation Parameters}
\tablewidth{0pt}
\tablehead{
Run & $\Omega_{\rm ce}/\omega_{\rm pe}$ & $m_{\rm i}/m_{\rm e}$ & $n_{\rm cr}/n_{\rm i}$ & $V_{\rm d,cr}/c$ & cell}
\startdata
A          & $3.26\times 10^{-2}$ & 10 & 1/20 & 0.1 & $4096\times 4096$\\
B          & $4.52\times 10^{-2}$ & 10 & 1/3  & 0.3 & $512 \times 512 $\\
C          & $3.26\times 10^{-2}$ & 10 & 1/20 & 0.2 & $2048\times 2048$\\
D          & $3.26\times 10^{-2}$ & 40 & 1/20 & 0.1 & $4096\times 4096$\\
\enddata
\tablecomments{Parameters of the simulation runs described in this paper. Listed are: 
the ratio of electron cyclotron frequency and electron plasma frequency $\Omega_{\rm ce}/\omega_{\rm pe}$, 
ion-electron mass ratio $m_{\rm i}/m_{\rm e}$, 
the density ratio of ambient ions and cosmic rays $n_{\rm cr}/n_{\rm i}$,
the velocity ratio of cosmic-ray drift and light speed $V_{\rm d,cr}/c$, the cell number}
\label{table1}
\end{deluxetable}

\end{document}